\documentclass{elsart}
\usepackage{epsfig}
\usepackage{natbib}

\def\url#1{#1} 

\begin{document}

\begin{frontmatter}

\title{The spectrum of HD46223.}

\author{Fr\'ed\'eric \snm Zagury}

\address{
\cty 02210 Saint R\'emy Blanzy, \cny France
\thanksref{email} }

   \thanks[email]{E-mail: fzagury@wanadoo.fr}

\received{May 2001}
\accepted{July 2000}
\communicated{L.L. Cowie}

 \begin{abstract}
The spectrum of HD46223 was established from the optical to the far UV and 
normalized by the spectrum of a non reddened star of same spectral 
type.
The resulting spectrum is separated into two components.
One is the direct starlight.
The second is an additional component of light scattered at 
small angles.
 
In the optical the spectrum is dominated by direct starlight which 
decreases exponentially due to the linear extinction $\propto e^{-2E(B-V)/\lambda}$.

Scattered light begins to be noticeable in the near-UV.
The near-UV rise of the scattered light is interrupted in the $2200\,\rm 
\AA$ bump region.
The wavelength dependence of the scattered light is established in the 
far-UV, where scattered light dominates the extinction curve. 
A $1/\lambda^4$ dependence is found, proving the presence of grains small 
compared to UV wavelengths.

The mathematic expressions of the different components mentionned above 
give a good fit to the extinction curve in the direction of the star.
On a mathematic standpoint the fit can be completed by a Lorentzien 
for the $2200\,\rm\AA$ bump region.
The physical interpretation of the bump may be more difficult to 
achieve since the paper shows the possibility that only
scattered light is extinguished in the bump region.

Consequences for the grain size distribution which is necessary 
to explain the different aspects of scattering in interstellar clouds, 
for the value of $R_V$, and problems raised by this interpretation of 
the spectrum of HD46223, are considered at the end of this paper.
   \end{abstract} 

\end{frontmatter}

 \section{Introduction} \label{intro}
This paper analyses the main features of the spectrum of HD46223 
at optical and UV wavelengths.
I pursue the analysis of the UV spectrum of the stars 
already initiated in two recent papers (\citet{uv1}, UV1 
hereafter, \citet{uv2}, UV2) and will follow an important conclusion of these papers:
the UV light we receive from a reddened star contains a non negligible proportion
of starlight scattered at very small angular distance from the star.

For a few stars with little reddening this conclusion was proved to be 
an alternative explanation for the departure of the UV 
extinction curve from the linear optical extinction (UV2). 
There are several reasons for which the UV spectrum of more heavily reddened stars is more 
difficult to interpret and therefore was not included in UV2.

The separation between scattered and direct starlight was natural for 
the stars presented in UV2.
The optical exponential decrease of the direct starlight was proved to 
extend in the near-UV, down to the bump region.
The scattered light, confined to the wavelength domain of highest 
optical depth, appeared as an excess of light superimposed on the tail 
of the exponential decrease.

With increasing reddenings, the separation is less obvious.
The exponential decrease becomes much steeper in the near-UV, making difficult 
its' precise determination from the UV spectrum alone.
If scattered light merges at 
longer wavelengths, the identification of 
the exponential decrease will be even harder.
To fix the slope of the linear extinction in regions of higher column density
we can adopt $E(B-V)$ from existing databases, 
but  a fit of the complete spectrum of the star, extending further in 
the optical, will be more reliable and convincing.

Optical wavelengths are necessary to understand how the 
scattered light component, important in the far-UV, extends in the 
near-UV and in the optical.
For the stars with little reddening (UV2) 
the scattered light predominantly occupies the far-UV side of the bump.
This is easily understandable since the extinction of starlight 
increases as the inverse of wavelength, hence increasing the number 
of photons available for scattering in the far-UV.
For larger values of $E(B-V)$, the scattering optical depth at 
constant $\lambda$ is increased due to the increase of column density.
The scattering component will be displaced towards the optical and the spectrum of a star 
restricted to the UV may not be wide enough to separate linear extinction 
and scattered light.

A small excess in the near-UV, above the exponential decrease, is perceived 
in the spectra presented in UV2 and can presumably be attributed to the 
extension of scattered light to longest wavelengths.
Although it remains within the noise level for small $E(B-V)$, as in UV2, 
it becomes a clear and independent feature with increasing 
reddening, for HD46223 for instance.
On Seaton's average extinction curve \citep{seaton79} it corresponds to the first 
split between $3\,\mu\rm m^{-1}$ and the bump of the extinction 
curve and the extension of the linear optical extinction.
The particular location of the split, at the junction between optical 
and UV wavelengths, further justifies to study 
extinction simultaneously in the UV and in the optical.

HD46223 is a O5e star with $E(B-V)\sim 0.5$.
Its' spectrum, from $8000\,\rm\AA$ to the far UV is established from 
the data presented in section~\ref{data}.

Section~\ref{redsp} gives the reduced spectrum 
of the star, defined in UV2 as the ratio of the star spectrum to the 
spectrum of a non reddened star of same spectral type.

The expected mathematic expressions of the scattered and of the 
direct starlight are established in sections~\ref{2comp} and \ref{expdec}.
Correction for linear extinction, which must be similar for direct 
and scattered light, leads to the determination of the wavelength 
dependence of the scattered light and to the fit of the reduced spectrum of 
HD46223 (section~\ref{tolambda}).

Section~\ref{dis} discusses the first implications and problems 
raised by the interpretation given in the preceding sections.
Some questions are deliberately left without answer.
They are treated in more detail in next papers.
\section{Data}\label{data}
\begin{figure*}[p]
\resizebox{!}{1.5\columnwidth}{\includegraphics{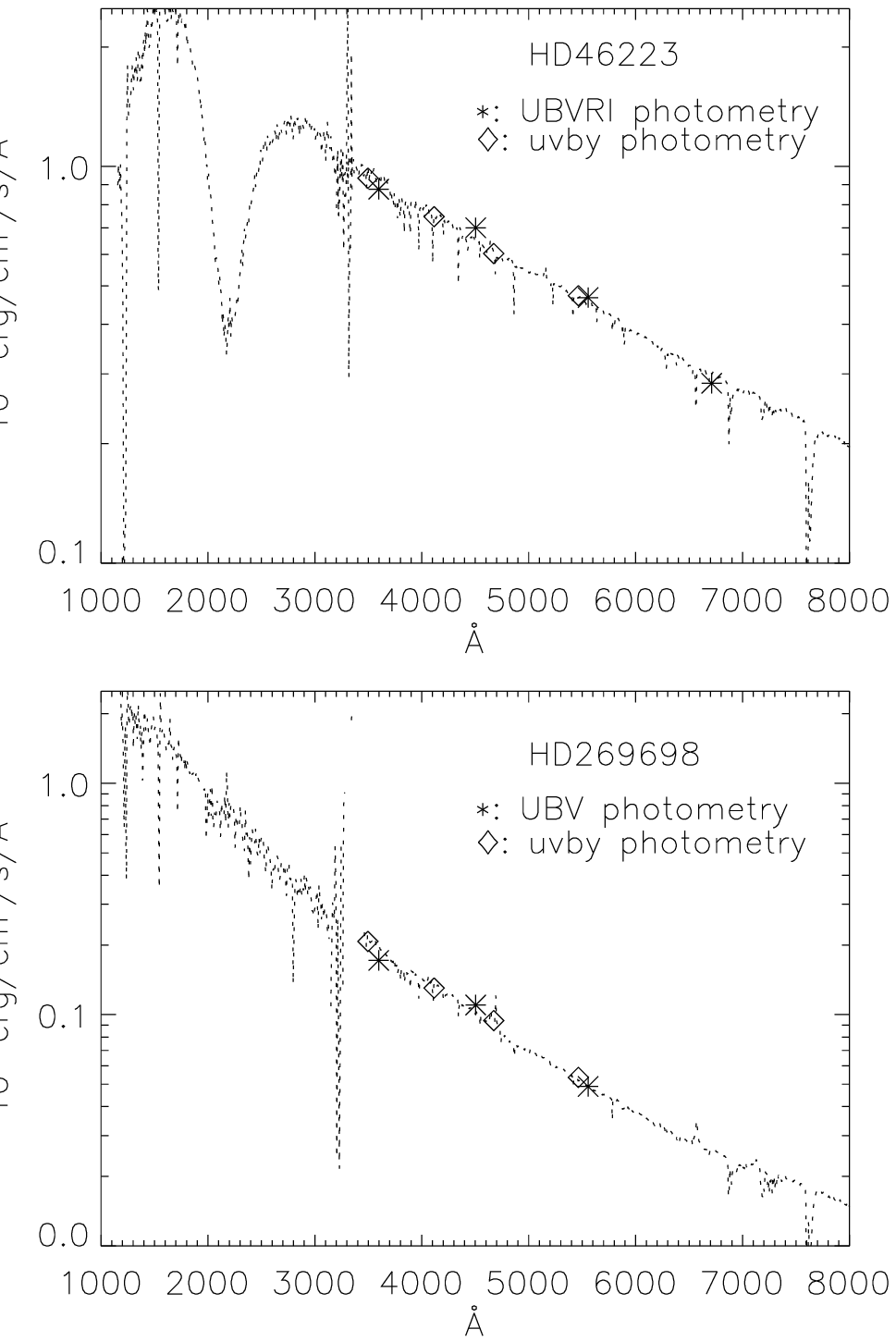}} 
\caption{Spectra of HD46223 (top) and of HD269698 (bottom).} 
\label{fig:pres}
\end{figure*}
    \begin{table}[tp]
       \[
    \begin{tabular}{|l|l|l|c|c|c|c|c|c|c|}
\hline
{\tiny name}& $\alpha_{1950}$ 
&$\delta_{1950}$&l&b&{\tiny sp. type}&{\tiny$ B$}
&{\tiny $B-V^{\,1}$}&
{\tiny $E(B-V)^{\,1}$}&{\tiny$ E(B-V)^{\,2}$}
   \\
\hline
{\tiny HD 46223}&{\tiny $06\,29\,29$}&{\tiny $+04\,51\,39$}
&206.44&-2.07&O5e&7.45&0.13&0.5&0.63 \\
\hline
{\tiny HD 269698}&{\tiny $05\,31\,50$}&{\tiny $-67\,40\,12$}
&277.84&-32.49&O5e&12.05&-0.22&0.1&0.1 \\
\hline
\end{tabular}    
    \]
\begin{list}{}{}
\item[$1$] $B-V$ from Simbad (\url{http://simbad.u-strasbg.fr}) and 
$(B-V)_0=-0.32$ from \citet{fitzgerald}
\item[$2$] $E(B-V)$ calculated from the fit of the extinction curve 
in the direction of HD46223.
\end{list}
\caption[]{}		
\label{tbl:star}
\end{table}
The data used in this paper are the UV and optical spectra of HD46223 and 
HD269698.

The UV spectra are extracted from the IUE database (\url{http://iuearc.vilspa.esa.es}).
The data reduction procedure I have used is described in UV1. 

The optical spectra are provided by J.F.~Le~Borgne (Le~Borgne et al. in preparation).
They belong to the stellar 
library accessible at: \url{http://webast.ast.obs-mip.fr/stelib/}.
HD46223 was observed at La Palma Observatory with a $1\,$m telescope.
The spectrum consists of $4$ sub-spectra with overlapping regions.
HD269698 was observed at Mount Stromblo Observatory with a $2.5\,$m telescope.
The spectrum consists of $6$ sub-spectra. 
The final resolution of the spectra is $\sim 1\,\rm\AA$ per pixel.
The reliability of the spectra was checked by the comparison with the 
Johnson  and the uvby photometry. 

\begin{figure*}[p]
\resizebox{!}{1.5\columnwidth}{\includegraphics{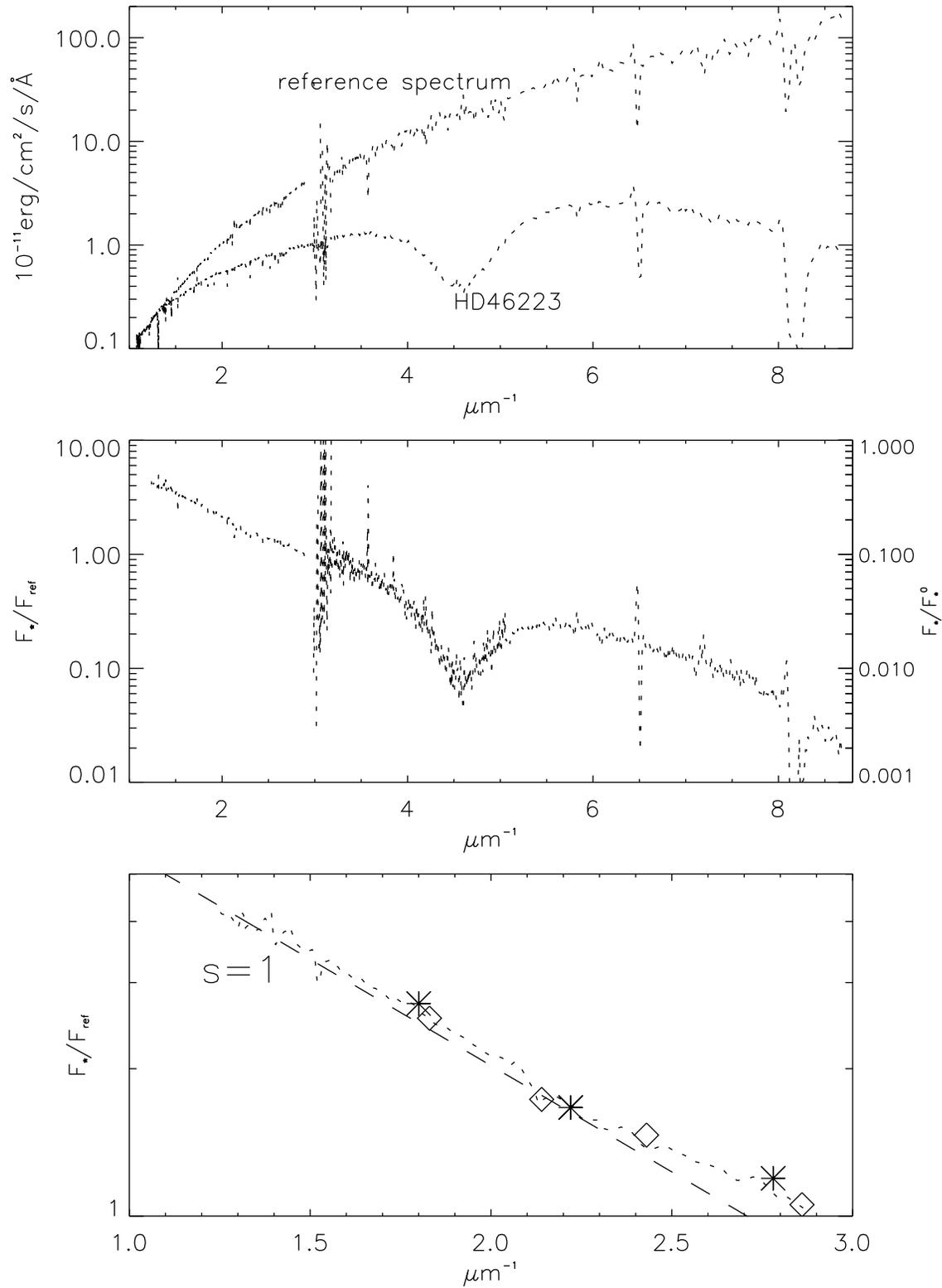}} 
\caption{\emph{top}: spectra of HD46223 and of the comparison spectrum 
used to establish the reduced spectrum of HD46223.
\emph{middle}: reduced spectrum of HD46223.
Right hand coordinates are absolute coordinates assuming $E(B-V)=0.5$ 
and $R_V=3$.
\emph{bottom}: reduced spectrum of HD46223 in the optical and 
underlying exponential $\propto e^{-1\mu \rm m/\lambda}$. UBV and ubvy 
photometry are also plotted.
} 
\label{fig:spred}
\end{figure*}

For both stars the UV and optical spectra adjust well (figure~\ref{fig:pres}).
No scaling factor was necessary to adjust the optical and UV parts of 
the spectra.
Longward of $1/\lambda=8\,\mu\rm m^{-1}$, in the neighborhood of 
Ly$\alpha$, the IUE calibration is poor (see \citet{fitzpatrick90} for a 
detailed discussion).
This is especially true for HD268698 whose
part of the spectrum after the Ly$\alpha$ absorption is outside 
the average slope indicated by the overall shape of the UV spectrum.
Consequently the reduced spectrum of HD46223 (bottom plot of 
figure~\ref{fig:spred}), presented in the next 
section, is lower than it should be for $1/\lambda>8.3\,\mu\rm m^{-1}$.
Since the possible error only concerns a small and particular part of the spectrum I 
chose not to make any correction.
In compensation the fit of the reduced spectrum of HD46223 presented 
in the paper is not asked to be accurate in the Ly$\alpha$ 
neighborhood. 

The properties of each star are summarized in table~\ref{tbl:star}.
Both are O5e stars. 
HD46223 is reddened and presents a bump at $2200\,\rm\AA$.
HD269698 does not have any bump, which indicates that the quantity of 
interstellar matter in front of the star is small, in conformity with 
the small value of $E(B-V)$ deduced from the magnitudes and the 
spectral type of the star.
\section{The reduced spectrum of HD46223} \label{redsp}
The reference spectrum used to compute the extinction curve in the 
direction of HD46223 is obtained by correcting the spectrum of 
HD269698 for the slight reddening in this direction.
The additional component of scattered light in the spectrum of HD269698 is considered to be 
negligible in agreement with the absence of a bump at $2200\,\rm \AA$.
Therefore the spectrum is multiplied by a factor 
$\propto e^{2E(B-V)/\lambda}=e^{0.2/\lambda}$ to correct the spectrum 
for the linear extinction which, according to UV2, extends to the UV.
The resulting spectrum is plotted on the top plot of 
figure~\ref{fig:spred}.

The middle plot of figure~\ref{fig:spred} shows the reduced spectrum 
of HD46223, defined up to a multiplicative constant.
The calibration of the reduced spectrum depends on $E(B-V)$, which is
determined from the slope of the optical part of the spectrum, and on 
the value of $R_V$ (see UV2).

A first approximation of the exponential decrease in the direction of HD46223 
is given by the exponential which best fits the 
optical part of the reduced spectrum (bottom plot of figure~\ref{fig:spred}).
The exponent, $1$, gives $E(B-V)=0.5$ in conformity with the value 
determined from the UBV photometry and the spectral type of HD46223.
On the same plot the ratios, in UBV and uvby photometries, of HD46223 
to the reference spectrum have also been represented.

As in UV2 an approximate absolute calibration of the reduced spectrum is 
estimated (middle plot, right y-axis),  assuming $E(B-V)=0.5$ and $R_V=3$.
\section{Direct and scattered light} \label{2comp}
As in UV2, the reduced spectrum of HD46223 will be 
separated into two parts.
The direct starlight refers to the photons 
which reach the observer directly from the star.
The scattered light is starlight re-introduced into the beam of the 
observation after one scattering at small angle.

In the optical, the extinction optical depth of the direct starlight is 
linear (UV2):
\begin{equation}
    \tau_{ext}=2E(B-V)[1/\lambda+(R_V-4)/2.2]
\label{eq:taulin}
\end{equation}
with $R_V=A_V/E(B-V)$, 
$\lambda$ in $\mu$m, and for central wavelengths of the $B$ and $V$ 
filters: $1/\lambda_B=2.27\,\mu\rm m^{-1}$, $1/\lambda_V=1.82\,\mu\rm m^{-1}$
\citep{cardelli}.

The direct starlight should be close to the straight line drawn on the 
middle and bottom plots of figure~\ref{fig:spred}.
The scattered light corresponds to the part of the spectrum above the 
exponential.
It represents nearly all the spectrum for $1/\lambda>5\,\mu\rm m^{-1}$.
\section{Mathematic expression of the reduced spectrum of HD46223} \label{expdec}
\subsection{Mathematic expression of the direct starlight} 
\label{matd}
The mathematic expression of the direct starlight in the reduced 
spectrum of HD46223 is $a_ne^{-\tau_{ext}^d}$.
$a_n$ is a constant (a reduced spectrum is defined up to 
a multiplicative constant), and $\tau_{ext}^d$ the wavelength dependent extinction optical 
depth of the interstellar matter between the star and us.
$a_n$ is $1$ if the reference (non-reddened) star is HD46223, 
corrected for reddening.

Moving toward the optical $\tau_{ext}^d$ is given by 
equation~\ref{eq:taulin} and $e^{-\tau_{ext}^d}$ is proportional to 
$e^{2E(B-V)/\lambda}$.
As a first approximation $E(B-V)=0.5$ can be adopted for HD46223.
\subsection{Mathematic expression of the scattered light} 
\label{mats}
Scattered light will be represented by a function $\tau_{sca}e^{-\tau_{ext}^s}$.
$\tau_{sca}$ can arbitrarily be defined as the 
ratio (scattered brightness)/(extinction along the path the photons 
have followed in interstellar space).  
This expression of the scattered light was proved in the single scattering 
case \citep{zagury99} with $\tau_{sca}=\omega \tau_{ext}$ and 
$\omega$ the albedo of the grains.

Since the interstellar path of the photons which reach the observer, 
whether they belong to scattered 
light or direct starlight, are similar, $\tau_{ext}^s$ must be 
equal or close to $\tau_{ext}^d$.

$\tau_{sca}$, related to the scattering cross-section of the grains, 
will be searched as a power law of $1/\lambda$: $\tau_{sca}\propto 
1/\lambda ^p$.
Theory predicts values of $p$ of order $1$ and a strong forward 
scattering phase function in case of large particles, $p\geq 4$ and an 
isotropic phase function for particles small compared to the 
wavelength. 
\section{Wavelength dependence of $\tau_{ext}^d$, $\tau_{ext}^s$, and 
$\tau_{sca}$ ouside the bump region} \label{tolambda}
\begin{figure*}[]
\resizebox{!}{1.5\columnwidth}{\includegraphics{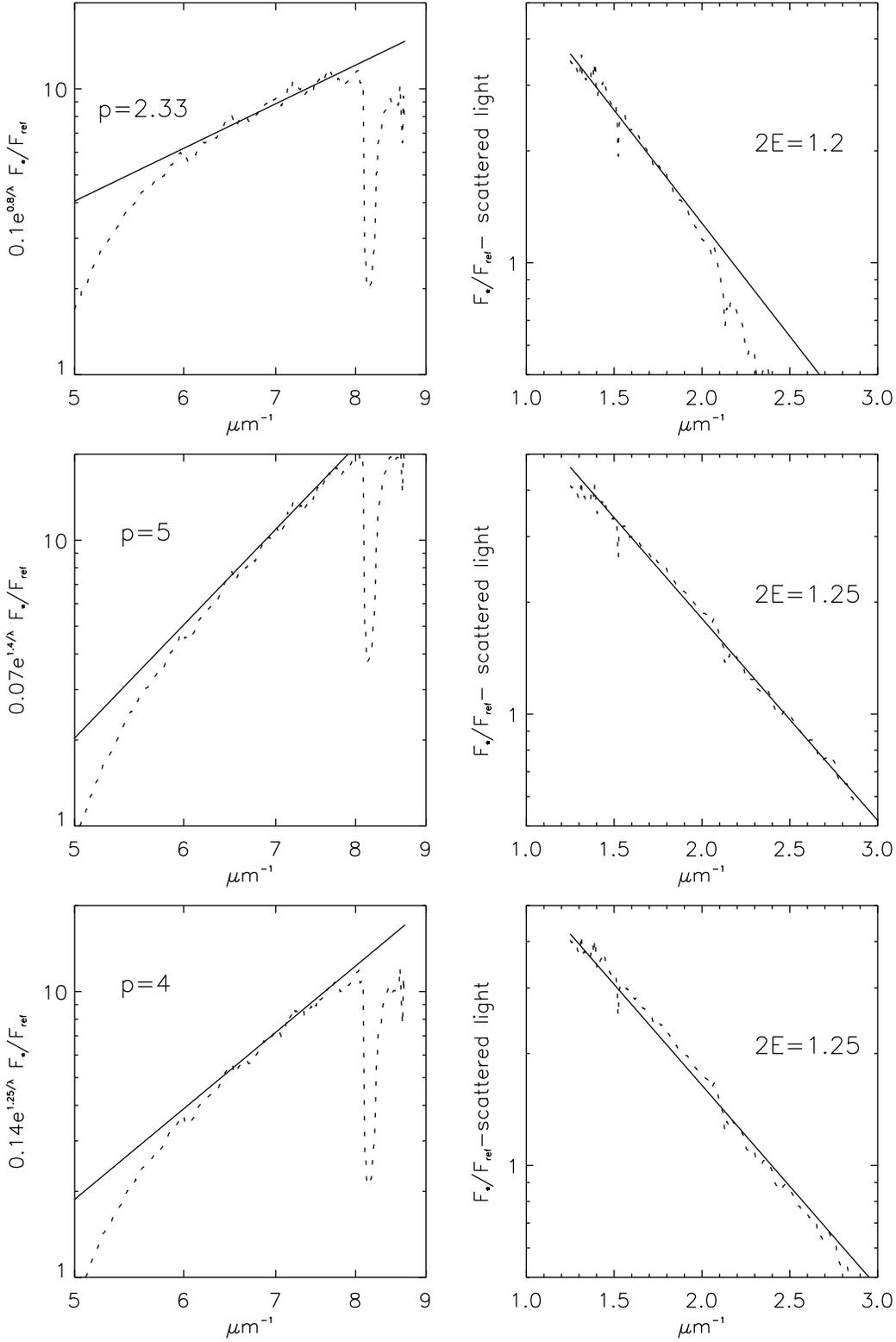}} 
\caption{Successive approximations to determine $\tau_{ext}^d$, 
$\tau_{sca}$ and $\tau_{ext}^s$ in the optical and in the far-UV.
Left hand plots are the successive approximations of the direct 
starlight in the optical, modelled by an exponential $e^{-2E/\lambda}$.
Right hand plots (log-log coordinates) will give the exponent $p$ of 
the scattering cross section for the scattered light.
The system stabilises at $2E=1.25$ and $p=4$.} 
\label{fig:sca}
\end{figure*}
Outside the bump region $\tau_{ext}^d$, $\tau_{sca}$ and 
$\tau_{ext}^s \sim \tau_{ext}^d$ will be determined using an iterative procedure.

The exponential decrease is first approximated by the exponential plotted on the 
bottom plot of figure~\ref{fig:spred}, resulting in: 
$\tau_{ext}^d=1/\lambda$, $2E=2E(B-V)=1$.

In the far-UV direct starlight can be discarded.
A first estimate of $p$  ($\tau_{sca}\propto 1/\lambda^p$) and of the 
scattered light is obtained 
by dividing the far-UV reduced spectrum by $e^{-\tau_{ext}^s}\sim 
e^{\tau_{ext}^d}$ (left and top plot of figure~\ref{fig:sca}).

The extension of the scattered light in the optical must then be substracted 
from the 
optical part of the spectrum (right and top plot of figure~\ref{fig:sca}).
A second exponential fit of the direct starlight in the optical, 
with exponent $2E$, is deduced (middle left plot of figure~\ref{fig:sca}), 
which in turn is used to divide the far-UV part of the spectrum (middle 
right plot of figure~\ref{fig:sca}).

\begin{figure*}[]
\resizebox{!}{1.5\columnwidth}{\includegraphics{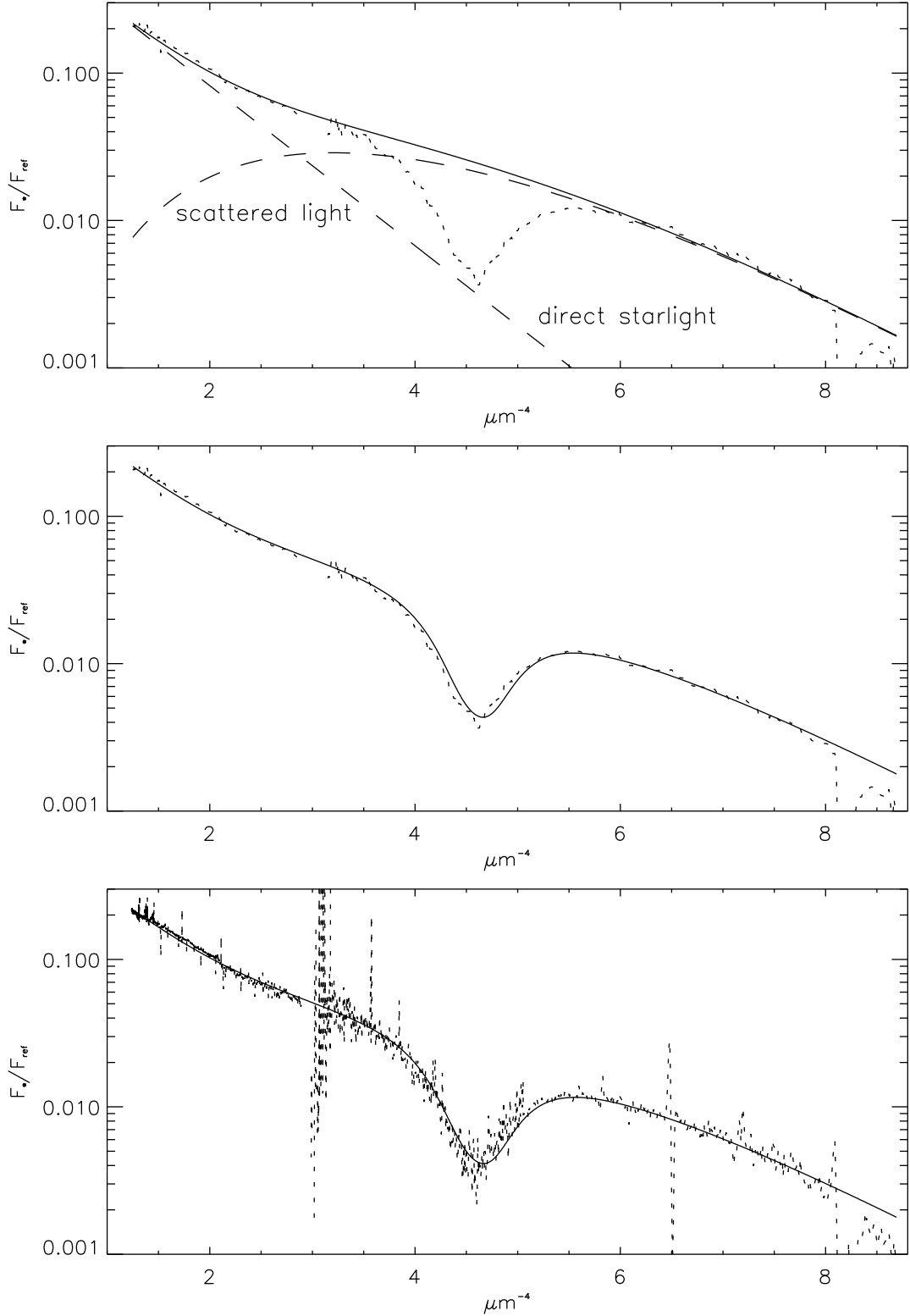}} 
\caption{\emph{top}: Optical and far-UV fit of the reduced spectrum 
of HD46223.
The two components, scattered and direct starlight are also separatly plotted.
\emph{middle}: The extinction curve in the direction of HD46223 
is fitted by the two components mentioned above and a Lorentz profile 
for the bump region.
The fit of the bump is a multiplicative component which can be equally 
applied to the fit of the top plot or restricted to the scattered 
light (more important in the bump region).
\emph{bottom}: Same plot as the middle one using the original,not 
smoothed, spectrum.} 
\label{fig:fit}
\end{figure*}

After two iterations the process is stabilized at $2E=1.25$ and $p=4$.
The resulting fit, for the optical and far-UV parts of the reduced 
spectrum of HD46223 is plotted top plot of figure~\ref{fig:fit}.
The mathematic expression of the fit:
\begin{equation}
   C(ax^4+1)e^{-2Ex}=C(0.015x^4+1)e^{-1.25x}
      \label{eq:fit}
\end{equation}¥
with $x=1/\lambda$. 
$C$ is an arbitrary constant which was fixed to $1$ in figure~\ref{fig:fit}.
The absolute spectrum is obtained for $C_{abs}=e^{-0.9E(R_V-4)}$ (UV2). 
For $R_V=3$, $C_{abs}=1.8$.

The fit of the bump region is not of real interest until the exact nature 
of the bump can be understood.
From a strictly mathematic point of view the bump can be fitted by a an 
exponential with exponent a
Lorentz \citep{savage75} or a Drude profile \citep{fitzpatrick90}.
In the middle and bottom plots of figure~\ref{fig:fit} a lorentzien fit
was adopted.
\section{Discussion} \label{dis}
\subsection{Far-UV and optical regions} \label{intfuv}
In the optical the reduced spectrum of HD46223 is essentially direct 
starlight extinguished with a linear in $1/\lambda$ extinction cross section.
The linear extinction of the direct starlight extends in the UV,
as it is predicted by the theory \citep{vandehulst} of the interaction 
between light and dust particles.

The slope of the reduced spectrum of HD46223 in the optical ($2E=1$) is  
smaller than the slope of the extinction of the direct 
starlight ($2E=1.25$).
This is due to the rise of the scattered light in the optical and 
towards the UV.

The scattered light is concentrated in the two bumps above the 
exponential decrease on each side of 
$1/\lambda_b=4.6\,\mu\rm m^{-1}$.
Compared to the spectrum of stars with little reddening presented in 
UV2 the singularity of the reduced spectrum of HD46223 is the 
importance of the near-UV bump clearly above the noise level.
For a given wavelength, and more specifically when moving towards the shortest 
wavelengths, the number of photons from HD46223 which are  
extinguished is larger than for the stars of UV2.
A larger number of photons are available for scattering.
In the spectral regions of low optical depth (towards the optical) 
scattering is enhanced, this explains the emergence of scattering in the near-UV.

Along with the increase of extinction and of scattered light in the optical the 
number of photons scattered in the far-UV is diminished.
This tendency was observed for the stars of UV2 (figure~6, 
bottom left plot of UV2).
In UV2 the maximum of the far-UV bump decreases with increasing reddening from $0.15$ for 
HD23480 ($E(B-V)=0.1$) to $\sim 0.02$ for HD62542 ($E(B-V)=0.4$) and 
HD200775 ($E(B-V)=0.5$).
For HD46223 it is of the same order as for HD200775, with a maximum of $\sim 0.02$ 
(middle plot of figure~\ref{fig:spred} or figure~\ref{fig:fit} with a scaling factor $C_{abs}\sim 1.8$).
When extinction is important, as it is in the direction of HD46223 in the far-UV,
it is dominated by absorption:
the increase of available photons for scattering is balanced in the 
far-UV by the increase of absorption.
Scattered light, as direct starlight, is extinguished by 
the linear extinction, more important here than for most of the stars presented in UV2.

At $\lambda_b$, the reduced spectrum of HD46223 is close to the exponential 
which represents the extinction of the direct starlight (top plot of 
figure~\ref{fig:fit}):
the near-UV rise of the scattered light is interrupted by the bump and 
the scattered light completely disappears at the bump position.

The $1/\lambda^4$ wavelength dependence of the scattered light proves
the presence of small (compared to UV wavelengths) interstellar grains.
Theory (\citet{vandehulst}, section~6.13, p.66) predicts a corresponding extinction cross 
section $\tau_{ext} \propto 1/\lambda$.

\subsection{The $2200\, \AA$ bump region} \label{intbump}
The $2200\,\rm\AA$ bump was fitted by a Lorentz function, as it is 
traditionally done.
The underlying standard interpretation considers the bump as an extinction feature due 
to a particular class of grains.
The mathematic expression of the bump can not held as a proof 
of this physical origin since different functions can reproduce 
the bump.
The standard interpretation of the bump has never received 
observational confirmations which would have identified a specific particle 
as responsible for extinction at $2200\,\rm\AA$.

The prolongation of the extinction of the direct 
starlight (middle plot of figure~\ref{fig:fit}) in the UV passes 
by or close to the lowest point of the bump.
This suggests that the process responsible for the bump affects 
only the scattered light, not the direct starlight.
The UV2 study of the reduced spectrum of stars with little reddening 
also indicates that direct starlight may not be affected by the bump.
For these stars there is no evidence of scattered light shortward of 
$1/\lambda_b$. 
The exponential decrease of the direct starlight is prolonged to 
$\lambda_b$, without any depression in the reduced spectrum (see for 
instance the case of HD23480 or of HD149757, figures~4 and 5 in UV2).

If it is confirmed that direct starlight is not specifically 
extinguished at $2200\,\rm\AA$, alternative explanations of the bump 
will have to be looked for.
\subsection{$E(B-V)$, $A_V$ and $R_V$.} \label{ebv}
The exact value of $E(B-V)$ in the direction of HD46223 is  half the slope of the 
extinction of the direct starlight (equation~\ref{eq:taulin}), $\sim 0.67$~mag.
This value is $25 \%$ larger than the one ($0.5$) deduced from the 
consideration of the optical part of the spectrum alone.

The exact value of the extinction coefficient $A_\lambda^0$ at 
wavelength $\lambda$ in the direction of the star will be retrieved 
from the fit of the direct starlight.
The error $\Delta A_V=A_V-A_V^0$ which is made when 
$A_V^0$ is confused with the observed value $A_V$ does not
depend on the absolute calibration of the reduced spectrum.
It is:
\begin{equation}
    \Delta A_V=-2.5\log(\frac{e^{-1/\lambda_V}}{e^{-1.25/\lambda_V}})
    =-0.27\,\mathrm {mag}
    \label{eq:av}
\end{equation}¥
$R_V$ is also modified.
If $E$, $E^0$ and $R_V$, $R_V^0$ are the observed and exact values 
of $E(B-V)$ and $R_V$:
\begin{equation}
    \Delta R_V=R_V-R_V^0=
    (\frac{E^0}{E}-1)R_V^0+\frac{\Delta A_V}{E}
    \label{eq:rv}
\end{equation}¥
In the present case:
\begin{equation}
    \Delta R_V=0.25R_V^0-0.21
    \label{eq:rvhd46223}
\end{equation}¥
Since $R_V^0$ is certainly larger than $1$, the observed value of $R_V$ is 
larger than $R_V^0$.
\subsection{Fit of the UV spectrum of the stars and the properties of 
interstellar grains} \label{disfit}
The decomposition of the spectrum of HD46223 was possible because of 
the large spectral coverage used to obtain a precise fit
of the extinction curve.
Acceptable fits to the UV part of the spectrum of HD46223, when extended to the 
optical, will proved to be inadequate to represent the optical spectrum.
The optical spectrum alone is not enough to determine accurately the 
linear extinction nor the UV part of the spectrum.

A large spectral coverage is also necessary to the determination of the exact value of $E(B-V)$.
If the value of $E(B-V)$ is not retrieved from the extinction curve it 
will be approximated, but with less precision, from $B-V$ and the 
spectral type of the star.

The number of parameters necessary to fit the extinction curve 
to a standard mathematic expression indicates the variability of the 
physical properties of the grains with direction of space.
In the standard conception of the extinction curve, the Fitzpatrick and 
Massa \citep{fitzpatrick90} decomposition requires five to six 
parameters to adjust the UV extinction curve in a given direction.
Two parameters are necessary for the $2200\,\rm \AA$ bump and two to 
three additional ones will fit the UV extinction curve outside the bump 
region.
Hence, great variability from region to region is thought to exist, 
which is explained by the existence of 
three type of interstellar grains.
The proportion of each type of grains depends on the direction which is 
considered.

Outside the bump region, the fit which is proposed in this paper for the 
reduced spectrum of HD46223, if it can be generalised to other directions, involves at 
the most one parameter.
This parameter measures the degree of scattered light which 
contaminates the spectrum of the star.
Since the importance of the far-UV bump is related to the reddening in the direction of the star 
(section~\ref{2comp}) a relationship between the parameter and $E(B-V)$ may exist.
If confirmed, the average properties of 
interstellar grains may be much more uniform than usually thought.

The light of HD46223, as for the stars in UV2, is exponentially 
extinguished with the same law in the optical and in the UV.
This is in conformity with the theory of extinction of light by dust 
particles which, in general, predicts a wavelength dependence of 
extinction as $1/\lambda$.
It also agrees with the principal result of UV1 which is that the light scattered by a 
nebula in the UV corresponds to a scattering cross-section linear in $1/\lambda$.

Considered together these results suggest that the average properties of 
interstellar grains are similar in all directions of space. 
\subsection{Grain size distribution} \label{disgr}
The $1/\lambda^4$ dependence found for the scattered light proves the 
presence of grains which are small compared to the UV wavelengths.
Small grains have properties radically different from the properties 
of the grains found in the study of the UV spectra of nebulae (UV1).
The wavelength dependence of the light scattered by a nebula observed 
on the side of the star was found to be 
linear in $1/\lambda$, excluding these grains as the 
possible carriers for the light scattered in the direction of the star.  

Small grains are not efficient forward scatterers:
small grains act as a small perturbation on the radiation source 
and produce a secondary wave which expands in all directions.
Scattering by small grains is not compatible with the strong forward scattering phase 
function necessary to explain the UV spectrum of the nebulae (UV1).
A strong forward scattering phase function, as it was evidenced in 
UV1, is characteristic of grains which are large compared to the wavelength.
Since scattering in the optical and in the UV have similar properties, 
the size distribution of the large grains must go above the 
dimensions of optical wavelength: 
grains larger than a few $\mu$m must exist in interstellar space.

The existence of two kinds of grains is necessary to explain the contamination of the 
spectrum of a star by scattered light and the UV spectrum of a nebula.
Large grains will explain the latter while small grains 
explain the UV spectrum of reddened stars.
It is a puzzling problem to understand how small grains with an 
isotropic phase function are more efficient in scattering light in the 
complete forward direction than large grains with a strong forward 
scattering phase function.
And why are the small grains not observed in the spectrum of the 
nebulae?
\section{Conclusion} \label{con}
Observations and studies of the spectrum of reddened stars have 
in most cases been conducted separately in the UV and in the optical. 
This purely formal discontinuity has favored separate 
interpretations of the UV and the optical extinction curves, and led 
to multi-components interstellar grain models.
In this paper, as in UV1 and UV2, I tried to show the continuity which exists 
between the optical and the UV, and how it may question the 
interpretation of the extinction curve.

The paper has focused on the spectrum of one star, HD46223, with 
moderate reddening, though larger than for the stars studied in UV2.
The spectrum was established from the optical (Le~Borgne's stellar 
library) to the far UV (IUE). 

The reduced spectrum of HD46223 was separated into two sub-spectra.
One for the starlight which directly reaches the observer.
One for starlight scattered at small angular distances from the star and 
re-injected into the beam of the observation.
A major difference of the spectrum of HD46223 with the spectrum of stars with smaller reddening 
is the emergence of scattered light above the noise level in the 
near-UV.
This is because enhanced extinction increases scattering at low 
optical depth.
But, the counterpart is a stronger absorption in the far-UV.

Both direct and scattered light cross the same media and must be 
extinguished in a similar way by the grains in the line of sight.
Each of these spectra dominates a specific wavelength range.
These properties were used to specify their dependence on wavelength.
The mathematic transcription of this physical interpretation of the 
spectrum proved to give an excellent fit of the reduced spectrum of HD46223.

The direct starlight is important in the optical and in the near UV.
It exponentially decreases with wavelength: linear extinction extinguishes 
starlight by factors of $150$ to $3000$ (for $E(B-V)=0.5$) 
from the bump region to the far-UV.
Linear extinction is present over all the spectrum.
It is also responsible for the ultimate far-UV decrease of the 
scattered light.

Scattered light becomes noticeable in the near UV, corresponding 
to the first depression between $2.5\,\mu\rm m^{-1}$ and $4.6\,\mu\rm m^{-1}$ observed 
on Seaton's average extinction curve.
The dependence of the scattering optical depth with wavelength is 
best identified in the far UV.
It varies as $1/\lambda^4$,
supporting the idea of scatterers of sizes small compared to the 
wavelength and a phase function close to isotropy.

The $2200\,\rm\AA$ bump corresponds to an interruption of the rise of the scattered light.
It is possible, as it is also indicated by the extinction curves, in directions of 
low reddening (UV2), that the bump affects the scattered light only, 
not the direct starlight.
This would question the standard interpretation of the bump.

The presence of scattered light extends to the optical.
Although it becomes a negligible part of the spectrum, it can be 
enough to substantially modify the true values of $E(B-V)$, $A_V$ and $R_V$.
$E(B-V)$ deduced from the magnitude of the star and its' 
spectral type underestimates the exact value of $E(B-V)$.
The observed $A_V$ is also larger than the true one because the 
introduction of scattered 
light gives the impression of lesser reddening.
In the case of HD46223 the observed $R_V$ is an overestimate of the 
true one.

The present paper indicates that it is possible to relate the physics 
in the extinction curve to the mathematical expression of the fit.
It also shows the necessity to consider spectra of 
reddened stars covering a large wavelength range.
Generalisation of the fit proposed in the paper may reduce the number 
of free parameters of the extinction curve, six in the 
decomposition of \citet{fitzpatrick90}.
Combined interpretation of the results presented in this paper, in 
UV1, and in UV2, leads to 
consider that the average properties of the interstellar grains may be much 
more uniform than previously thought.

This interpretation of the extinction curve also raises problems.
The difference in scattering cross section, with a $1/\lambda^4$ and a 
$1/\lambda$ dependence, for the scattered light which contaminates 
the spectrum of a star and for the light scattered by a nebula, 
distinguishes two types of grains. 
How can small grains with a near isotropic phase function be 
observed in the spectrum of a star, hence at a scattering angle close 
to $0$, and not be seen in a nebula at close angular distance from the star? 

{}

\end{document}